\documentclass[11pt,a4paper]{article}
\usepackage{a4wide}
\usepackage{amsmath}
\makeatletter
\@addtoreset{equation}{section}
\makeatother

 \def\unit{\hbox to 3.3pt{\hskip1.3pt \vrule height 7pt width .4pt \hskip.7pt
\vrule height 7.85pt width .4pt \kern-2.4pt
\hrulefill \kern-3pt
\raise 4pt\hbox{\char'40}}}

\def\half{{\textstyle {1 \over 2}}}

\def\Dpartial{{\cal D}}
\def\ap#1{\alpha^{\prime\,#1}}

\def\slash{\llap /}
\def\slashed#1{#1\!\slash\,}

\def\makeatletter{\catcode`\@=11}
\makeatletter
\def\mathbox#1{\hbox{$\m@th#1$}}%
\def\math@ccstyles#1#2#3#4#5#6#7{{\leavevmode
      \setbox0\mathbox{#6#7}%
      \setbox2\mathbox{#4#5}%
      \dimen@ #3%
      \baselineskip\z@\lineskiplimit#1\lineskip\z@
      \vbox{\ialign{##\crcr
             \hfil \kern #2\box2 \hfil\crcr
             \noalign{\kern\dimen@}%
             \hfil\box0\hfil\crcr}}}}
\def\mathaccstyles{\math@ccstyles\maxdimen}
\def\maththroughstyles{\math@ccstyles{-\maxdimen}}
\def\unitmatrixDT%
 {\maththroughstyles{.45\ht0}\z@\displaystyle {\mathchar"006C}\displaystyle 1}

\newcommand{\bea}{\begin{eqnarray}}
\newcommand{\eea}{\end{eqnarray}}
\newcommand{\nn}{\nonumber}

\newcommand{\beann}{\begin{eqnarray*}}
\newcommand{\eeann}{\end{eqnarray*}}

\newcommand{\pd}{\partial}      
\renewcommand{\L}{\mathcal{L}}  
\newcommand{\D}{\mathcal{D}}    
\newcommand{\Tr}{{\rm Tr}}      

\newcommand{\g}{\gamma}
\renewcommand{\d}{\delta}
\newcommand{\e}{\epsilon}

\renewcommand{\l}{\lambda}

\begin{document}


\pagestyle{empty}
\begin{flushright}
\small
UG-02/39\\
{\bf hep-th/0205150}\\
May, 2002
\normalsize
\end{flushright}

\begin{center}


\vspace{.7cm}

{\Large {\bf Supersymmetric Yang-Mills theory}}
\vspace{0.5cm}
{\Large {\bf  at order $\ap{3}$}}

\vspace{.7cm}


 A.~Collinucci\footnote{\tt a.collinucci@phys.rug.nl}, M.~de
 Roo\footnote{\tt m.de.roo@phys.rug.nl}
 and M.G.C.~Eenink\footnote{\tt m.g.c.eenink@phys.rug.nl}

\vskip 0.4truecm

\begin{center}
 Institute for Theoretical Physics\\
   Nijenborgh 4, 9747 AG Groningen\\
     The Netherlands\\
\end{center}
\vskip 1.0cm

{\bf Abstract}

\end{center}

\begin{quotation}

We construct the order $\ap{3}$ terms in the supersymmetric
 Yang-Mills action in ten dimensions for an arbitrary gauge
 group. The result can be expressed in terms of the structure
 constants of the Yang-Mills group, and is therefore independent
 of abelian factors. The $\ap{3}$
 invariant obtained here is independent of the $\ap{2}$ invariant,
 and we argue that additional superinvariants will occur at
 all odd orders of $\ap{}$.

\end{quotation}

\newpage

\pagestyle{plain}

\setcounter{section}{0}
\section{Introduction\label{Intro}}

The abelian Born-Infeld action provides us with an
 effective theory, which reproduces
 to all orders in $\alpha'$ the tree level scattering
 amplitudes of massless modes of open strings that end on a single D-brane,
 with the assumption that the fields vary
 slowly \cite{Tseyt}. As was recently shown in \cite{Bilal},
 this assumption implies that gravitational effects are large.
 Small
 derivatives imply that the fields stay large over a vast region, and
 an estimate of the total energy and the corresponding volume indicates
 that under gravitational forces such a system would collapse to a
 black hole.
 To avoid this, fields have to fall off
 over a short distance, making derivatives large. Physically it is
 hard to make sense of the Born-Infeld action in string theory, where
 gravitational forces are implied by the presence of closed strings.

When $n$ D-branes coincide, the gauge group is enhanced to $U(n)$
 \cite{Witten}, making the task of writing an effective action much
 more complicated. Now there is an additional, practical, argument
 for including derivative terms. A constant field strength is  not
 a gauge-invariant concept, and one has to take into account
 that $[\Dpartial,\Dpartial]F = [F,F]$.
 So, if we were to neglect derivatives of fields, we would also
 have to neglect  commutators of field strengths, which
 amounts to going back to the abelian situation.

In the abelian case the complete supersymmetric action for slowly varying
 fields is known \cite{Aga}. According to the argument of \cite{Bilal},
 this does not mean that derivative terms are necessarily small.
 Contributions to the abelian action involving derivatives
 have been obtained in \cite{AndTs,Wyll} from string amplitude
 calculations. Under supersymmetry
 these terms should form an invariant, which is independent of the
 Born-Infeld action.
 In the nonabelian case these invariants are no longer independent
 because of $[\Dpartial,\Dpartial]F = [F,F]$.
 A straightforward approach to the Yang-Mills case is then to
 compute tree-level string scattering amplitudes and calculate the
 corresponding effective action. This method has been applied
 to the string four-point function (see \cite{Schwarz} and
 references therein) and has yielded complete
 results for orders $\ap{2}$ \cite{BBRS}, and
 partial results for order $\ap{3}$ and $\ap{4}$ \cite{Bilal}. For
 instance, at order $\ap{3}$  terms of the form $(\Dpartial F)^2F^2$ and
 $F\Dpartial F\chi\g \Dpartial\Dpartial\chi$ (plus terms that are quartic in
 $\chi$, which we will
 not deal with in this paper) have been computed in this way.
 This leaves the $F^5$ and  $F^3\chi \g
 \Dpartial \chi$ terms to be determined.

Another approach consists in calculating the deformations allowed
 by supersymmetry of the $d=10$ super Yang-Mills theory.
 In \cite{BRS,goteborg} this idea is
 put to the test up to order $\ap{2}$.
 One finds that $\ap{}$ terms can be eliminated via field
 redefinitions, and
 the $\ap{2}$ terms match string theory predictions. In the
 calculation of \cite{BRS} a significant simplification is reached
 by the assumption that only symmetric traces of the Yang-Mills
 generators appear. A superspace calculation by \cite{goteborg}
 yields a result to all orders in the fermions,
 where all Yang-Mills indices enter symmetrically.
 At $\ap{3}$ a symmetric single trace
 is not possible, since the symmetric trace of $F^5$ vanishes.
 However, the string theoretical calculations
 performed in \cite{kita} show that terms of the form
 $F^5$ and $(\Dpartial F)^2F^2$ are needed.

Recently, two calculations of the bosonic $\ap{3}$ terms have
 been performed. In \cite{RSTZ} the one-loop five-point amplitude
 is calculated in $N=4$ super-Yang-Mills theory in four dimensions.
 This leads to an effective
 $\ap{3}$ action that reproduces this amplitude, and, assuming that
 supersymmetry uniquely determines such an action, this $N=4$ $d=4$
 result should then correspond (although not uniquely) to the ten
 dimensional effective action. In \cite{KS1} deformations of the $d=10$
 Yang-Mills theory that preserve a BPS solution to the equations of motion
 are studied. This method also yields an effective action at $\ap{3}$,
 now directly in $d=10$.
 Although \cite{RSTZ} and \cite{KS1}
 find the same $(\Dpartial F)^2F^2$ terms, they disagree on the $F^5$
 contributions.

We obtain in this paper the $\ap{3}$ terms
 in the effective action,
 including the terms bilinear in the fermions, by imposing supersymmetry
 to order $\ap{3}$. The result agrees with \cite{kita,RSTZ,KS1,Bilal}
 for the bosonic terms with
 derivatives, and with \cite{KS1} for the bosonic $F^5$ terms.
 The group structure of the action and transformation rules that
 we obtain can be expressed completely in terms of the structure
 constants. This implies that the result vanishes in the abelian case,
 and also that it is trivially invariant under nonlinear supersymmetry
 transformations, which act only on a $U(1)$ factor in the gauge group.

This paper is organized as follows. In Section \ref{Method} we
 explain our calculational method, showing, as an example,
 that no effective action at order $\ap{}$ is needed.
 The result at order $\ap{3}$  is presented
 in Section \ref{Result}. In Section \ref{Zeta} we construct
 the $\ap{}$-expansion of the string four-point function,
 and discuss consequences of this expansion for the
 effective action at higher orders in $\ap{}$. In particular, we will
 argue that al all odd orders in $\ap{}$ a new, independent, superinvariant
 begins. We present
 our conclusions in Section \ref{Discussion}.

\section{Constructing the order $\ap{3}$ action} \label{Method}

First we review the $d=10$ $N=1$ supersymmetric
 Yang-Mills theory in order to set the stage for our 
 calculations. The Lagrangian is given by\footnote{Our conventions for the 
 $\g$-matrices follow \cite{Proey}. For the gauge fields, they are
 presented in Appendix \ref{Conventions}. We will always write
 spacetime indices as lower indices.}:
\begin{equation}
\label{SYMLagr}
  \L_{YM} \;=\; -\frac{1}{g^2}
  \Tr\, \{ -\tfrac{1}{4}F_{ab}F_{ab}+\tfrac{1}{2}\bar{\chi}\slashed{\D}\chi \}
\,.
\end{equation}
$g$ is the Yang-Mills coupling constant; it has mass
 dimension $-3$. The gauge field $A_a$ and the derivatives  $\D$ and $\pd$
 have dimension +1, the
 gaugino $\chi$ dimension +3/2.
 From now on we will drop the factor of $1/g^2$ for notational
 clarity, the dimension of the remaining Lagrangian then equals +4.

Variation of this action gives $\d\L_{YM} = -\Tr\left\{(\D_a
 F_{ab}-\bar{\chi}\g_b\chi)\d
 A_b+\d\bar{\chi}\slashed{\D}\chi\right\}$, from which one obtains
 the equations of motion:
\bea
  0 &=& \D_a F_{ab}^A - \tfrac{1}{2} f^{ABC}\bar{\chi}^B\g_{b}\chi^C\,,\\
  0 &=& \slashed{\D}\chi^A \,.
\eea
$\mathcal{L}_{YM}$ is invariant under the following supersymmetry
 transformations:
\bea
\label{trans0A}
  \d_\e A_a &=& \bar{\e}\g_a\chi, \\
\label{trans0chi}
  \d_\e \chi &=& \tfrac{1}{2} F_{ab} \g_{ab}\,\e,
\eea
where $\e$ is a constant Majorana-Weyl spinor of dimension +1/2. As
 is well known, the supersymmetry algebra only closes on-shell and
 involves a field-dependent gauge transformation:
\bea
  {[\d_{\e_1},\d_{\e_2}]} A_a &=&
    2\bar{\e_1}{\pd}\slash\e_2
   A_a - \D_a \left(2\bar{\e_1}\slashed{A}\e_2\right),\nonumber\\
\label{algebra}  {[\d_{\e_1},\d_{\e_2}]} \chi &=&
 2\bar{\e_1}\slashed{\D}\e_2\chi -\left(
 \tfrac{7}{8}\bar{\e_1}\g_a\e_2\g_a -\tfrac{1}{5!16}\bar{\e_1}
    \g_{a_1\cdots a_5}\e_2\g_{a_1\ldots a_5}\right)\slashed{\D}\chi.
\eea
Before moving on to the actual $\ap{3}$ corrections
 to (\ref{SYMLagr}) we will first discuss our method.

Consider a general Lagrangian $\L_0[\phi]$
 that possesses a symmetry, with infinitesimal transformations
 $\delta_0\phi$.
 If $\L = \L_0 + \l\L_1$, where $\l$ is some expansion parameter,
 then the
 variation of $\L_1$ due to $\delta_0\phi$ generically yields terms that,
 to preserve the symmetry,
 should be cancelled by an $\l$ variation of
 $\phi$ in $\L_0$.
 Cancellation occurs if and only if
 the variation of $\L_1$ is proportional to the order $\l^0$
 equations of motion. The Lagrangian $\L$ one obtains in this way is
 uniquely defined up to total derivatives and field redefinitions.
 A field redefinition $\phi\to \phi + \lambda\Delta\phi$ gives rise
 to order $\lambda$ terms of the form
\begin{equation}
\label{fieldredef}
     \lambda\Delta\phi\, \frac{\d\L_0}{\d\phi_i}\,,
\end{equation}
i.e., is proportional to the
 order $\l^0$ equations of motion. Therefore, any term in  $\L_1$
 of the form (\ref{fieldredef}) can be eliminated by a field
 redefinition. We will choose our  $\ap{3}$ action such that no explicit
 terms of the form (\ref{fieldredef}) appear.

Let us illustrate how this works by considering  order $\ap{1}$.
 From this point on we discard any terms of higher than quadratic
 order in the fermions.
 Since $\ap{}$ has mass dimension $-2$ we can write down terms that
 have dimension $+6$. These terms must be Lorentz and gauge
 invariant. We also take only terms with a single trace over the
 generators $T^A$ since we want to make contact with a string
 theory tree level effective action. Possible terms are:
\renewcommand{\arraystretch}{1.3}
\[
  \begin{array}{lll}
    (1)\;\; & \Tr\, T^AT^BT^C    & F_{ab}{}^A F_{bc}{}^B F_{ca}{}^C, \\
    (2)\;\; & \Tr\, T^AT^BT^C    & \D_aF_{ab}{}^A \,\bar{\chi}^B\g_b\chi^C, \\
    (3)\;\; & \Tr\, T^AT^BT^C  & \D_aF_{bc}{}^A \,
      \bar{\chi}^B\g_{abc}\chi^C, \\
    (4)\;\; & \Tr\, T^A[T^B,T^C]    & F_{ab}{}^A \,\bar{\chi}^B\g_a\D_b\chi^C,
\\
    (5)\;\; & \Tr\, T^A\{T^B,T^C\}  & F_{ab}{}^A \,\bar{\chi}^B\g_a\D_b\chi^C,
\\
    (6)\;\; & \Tr\, T^A[T^B,T^C]    & F_{ab}{}^A \,\bar{\chi}^B\g_{ab}
      \slashed{\D}\chi^C, \\
    (7)\;\; & \Tr\, T^A\{T^B,T^C\}  & F_{ab}{}^A \,\bar{\chi}^B\g_{ab}
      \slashed{\D}\chi^C\,. \\
  \end{array}
\]
\renewcommand{\arraystretch}{1}
\noindent In choosing these terms we put no restriction on
 the group structure other than the cyclic property of the trace.
 We do not need to take
 terms with more than one derivative: it is not difficult to
 convince oneself that such terms always contain
 $[\D,\D]$ and/or lowest order equations of motion.
 We see that (3) vanishes due to the
 Bianchi identity. Furthermore, (2), (6) and (7) are proportional
 to the order $\ap{0}$ field equations, so we do not allow
 them in the Lagrangian. Since $\L$ is only defined up to total
 derivatives we also consider all of these:
\renewcommand{\arraystretch}{1.3}
\[
  \begin{array}{llcl}
    \Tr\, T^AT^BT^C   & \pd_a ( F_{ab}{}^A\,\bar{\chi}^B\g_b\chi^C )     &=&
(2)-(4),\\
    \Tr\, T^AT^BT^C & \pd_a ( F_{bc}{}^A\,\bar{\chi}^B\g_{abc}\chi^C ) &=& (3)+
(7)-2\times(5).
  \end{array}
\]
\renewcommand{\arraystretch}{1}
\noindent So we see that we also need not include (4) and (5) since they
 can be rewritten as
 a total derivative and terms that can be cancelled by a field redefinition.
 This analysis leaves only the term (1).

We now show that the remaining
 term $f^{ABC}F_{ab}{}^AF_{bc}{}^BF_{ca}{}^C$ is not allowed by
 supersymmetry. Varying this term with the
 transformation rule (\ref{trans0A})
 gives:
\[
  6f^{ABC}F_{ac}{}^AF_{cb}{}^B\,\bar{\e}\g_a\D_b\chi^C\,.
\]
We adopt the rule that any derivative on $\chi$ in a variation
 is partially integrated to act on the bosonic fields - except
 in the situation where this derivative takes on the form of the
 order $\ap{0}$ equation of motion $\slashed{\D}\chi^A$. This rule
 leads to
\begin{equation}
\label{var1}
  3f^{ABC}\D_aF_{bc}{}^AF_{bc}{}^B
  \bar{\e}\g_a\chi^C+6f^{ABC}\D_aF_{ab}{}^AF_{bc}{}^B\bar{\e}\g_c\chi^C\,.
\end{equation}
The second term in (\ref{var1}) 
 contains the $A_a$ equation of motion, and can therefore
 be cancelled by an order $\ap{}$ transformation, 
 while the first term cannot.
 Therefore term (1) does not allow supersymmetrization;
 the only terms
 allowed by supersymmetry at order $\ap{}$ can
 be eliminated by a field redefinition.

In the present case we can see by inspection that the first term in
 (\ref{var1}) cannot be rewritten as a total derivative plus terms
 containing equations of motion. In a more complicated situation one
 would parametrize all possible total derivatives, which lead to the
 same structures as those in (\ref{var1}) to verify this fact.

So our method comes down to the following: first we write down an action
 involving all possible terms that are independent up to partial
 integrations. To this we add all possible total derivatives,
 and use these to reduce the starting point to a
 minimal number of terms. This results in the Ansatz
 for the effective action, in which each term gets an arbitrary
 coefficient to be determined later on.
 We then vary the Ansatz with the lowest order variations of $A$ and $\chi$.

To this variation we add all possible total derivatives, which lead
 to contributions having the same structure as the variations.
 These also have arbitrary coefficients. All terms proportional to
 lowest order equations of motion of $A$ and $\chi$ are saved for later
 use in determining the new transformation rules. After eliminating all
 remaining derivatives on $\chi$ by partial integrations, the rest
 has to vanish, and this gives rise to linear equations between the unknown
 coefficients. Note that the fact that all variations are ultimately written
 without derivatives on $\chi$ implies that the total derivatives that we
 add to the variation must give rise to a lowest order fermion equation of
 motion - otherwise the partial integration away from $\chi$ just reproduces
 the original total derivative, and the term does not influence the
 calculation.
 An important part of the calculation is to rewrite the
 remaining terms such that the minimal number of independent structures
 is left. This is done by using Bianchi identities
 for $\Dpartial F$, $\Dpartial\Dpartial F$, etc.,
 and by ordering the field strengths. Each independent structure gives
 rise to an equation between the coefficients.
 If these equations have nontrivial solutions then these
 correspond to supersymmetric actions.

In the case of the $\ap{3}$ modification to the Yang-Mills action the
 number of terms at intermediate stages of the calculation
 reaches $10^4$. Therefore, the required
 algebraic manipulations, such as obtaining the variation of the Ansatz,
 working out products of $\gamma$-matrices,  partial integrations,
 the use of Bianchi identities, are all done by computer.

\section{SYM at order $\ap{3}$\label{Result}}

We saw that at order $\ap{}$ there are no nontrivial
 modifications to the supersymmetric action (\ref{SYMLagr}).
 At order $\ap{2}$ there are nontrivial corrections to the super
 Yang-Mills Lagrangian  and supersymmetry
 transformation rules \cite{BRS,goteborg}.
 However, in the iterative procedure these
 terms cannot contribute to the order $\ap{3}$ variations,
 precisely because there are no order $\ap{}$ terms in the
 transformation rules. This means that at $\ap{3}$ the analysis
 follows the outline given in the previous section.

However, there is one complication. At order $\ap{3}$ we have to go
 through a two-step procedure, since in the Ansatz we have not only terms
 with five fields, i.e., $F^5$ and the corresponding terms involving fermions,
 but also terms  with four fields, such as $(\Dpartial F)^2F^2$ with
 fermionic
 partners.
 In this case the analysis, both in determining the
 Ansatz and in cancelling the variation, has to start at the higher-derivative
 terms. The reason is that
 the higher-derivative terms produce terms with less derivatives
 because of $[\Dpartial,\Dpartial]F=[F,F]$
 and $[\Dpartial,\Dpartial]\chi=[F,\chi]$.
 It is easily seen that all terms with four derivatives
 and two $F$'s, and their fermionic partners, can be eliminated by field
 redefinitions.

The leading terms in this analysis are therefore the higher-derivative
 terms $(\Dpartial F)^2F^2$ and partners. As we mentioned before, the Ansatz is
 not unique. We found that the bosonic part of the Ansatz must contain
 13 terms (in agreement with \cite{KS1}),
 and one may choose for instance to have only $(\Dpartial F)^2F^2$
 terms and no $F^5$ terms
 \cite{KS2}. However, for the terms involving fermions
 the partners of $F^5$ cannot all be eliminated. We have chosen
 for the bosonic part of our
 Ansatz the 13 terms in the starting point of \cite{KS1}.
 Our Ansatz then contains 13 bosonic terms, and 110
 terms involving fermions:  $7+18$ terms of the form $(\Dpartial F)^2F^2$
 and fermionic partners, and $6+92$ of type $F^5$ with partners.

After simplifying the resulting variations there remain 128 linear
 equations from the sector with four fields, and 320 equations from the sector
 with five fields. These equations must be solved
 for the 123 coefficients from the Ansatz and the 182 coefficients
 that parametrize total derivatives having the same structure as
 the variations (see Section \ref{Method}).

The result is that
 there is {\em one} unique deformation of $d=10$, $N=1$
 supersymmetric Yang-Mills theory at order $\ap{3}$, up to a single
 multiplicative constant, which according to string theory 
 equals $\zeta(3)/2$.
 In one particular parametrization, the
 result is:
\begin{eqnarray}
{\cal L}_3 &=& f^{XYZ}f^{VWZ}\bigg[2\,F_{ab}{}^{X}F_{cd}{}^{W}
\Dpartial_eF_{bc}{}^{V}\Dpartial_eF_{ad}{}^{Y}
-2\,F_{ab}{}^{X}F_{ac}{}^{W}
\Dpartial_dF_{be}{}^{V}\Dpartial_dF_{ce}{}^{Y}
\nonumber\\
&&\qquad +F_{ab}{}^{X}F_{cd}{}^{W}
\Dpartial_eF_{ab}{}^{V}\Dpartial_eF_{cd}{}^{Y}
\nonumber\\
&&\qquad -4\,F_{ab}{}^{W}
\Dpartial_cF_{bd}{}^{Y}\bar\chi^{X}\gamma_{a}\,\Dpartial_d\Dpartial_c\chi^{V}
-4\,F_{ab}{}^{W}
\Dpartial_cF_{bd}{}^{Y}\bar\chi^{X}\gamma_{d}\,\Dpartial_a\Dpartial_c\chi^{V}
\nonumber\\
&&\qquad +2\,F_{ab}{}^{W}
\Dpartial_cF_{de}{}^{Y}\bar\chi^{X}\gamma_{ade}\,\Dpartial_b\Dpartial_c\chi^{V}
+2\,F_{ab}{}^{W}
\Dpartial_cF_{de}{}^{Y}\bar\chi^{X}\gamma_{abd}\,\Dpartial_e\Dpartial_c\chi^{V}
\bigg]\ +
\nonumber
\end{eqnarray}
\begin{eqnarray}
&& +\ f^{XYZ}f^{UVW}f^{TUX}\bigg[4\,
  F_{ab}{}^{Y}F_{cd}{}^{Z}F_{ac}{}^{V}F_{be}{}^{W}F_{de}{}^{T}
+2\,F_{ab}{}^{Y}F_{cd}{}^{Z}F_{ab}{}^{V}F_{ce}{}^{W}F_{de}{}^{T}
\nonumber\\
&&\qquad -11\,F_{ab}{}^{Y}F_{cd}{}^{Z}F_{cd}{}^{V}\bar\chi^{T}\gamma_{a}\,
\Dpartial_b\chi^{W}
+22\,F_{ab}{}^{Y}F_{cd}{}^{Z}F_{ac}{}^{V}\bar\chi^{T}\gamma_{b}\,
\Dpartial_d\chi^{W}
\nonumber\\
&&\qquad +18\,F_{ab}{}^{Y}F_{cd}{}^{V}F_{ac}{}^{W}\bar\chi^{T}\gamma_{b}\,
\Dpartial_d\chi^{Z}
+12\,F_{ab}{}^{T}F_{cd}{}^{Y}F_{ac}{}^{V}\bar\chi^{Z}\gamma_{b}\,
\Dpartial_d\chi^{W}
\nonumber\\
&&\qquad +28\,F_{ab}{}^{T}F_{cd}{}^{Y}F_{ac}{}^{V}\bar\chi^{W}\gamma_{b}\,
\Dpartial_d\chi^{Z}
-24\,F_{ab}{}^{Y}F_{cd}{}^{V}F_{ac}{}^{T}\bar\chi^{W}\gamma_{b}\,
\Dpartial_d\chi^{Z}
\nonumber\\
&&\label{L3}\qquad +8\,F_{ab}{}^{T}F_{cd}{}^{Y}F_{ac}{}^{Z}\bar\chi^{V}\gamma_
{b}\,
\Dpartial_d\chi^{W}
-12\,F_{ab}{}^{T}F_{ac}{}^{Y}
\Dpartial_bF_{cd}{}^{V}\bar\chi^{Z}\gamma_{d}\,\bar\chi^{W}
\\
&&\qquad -8\,F_{ab}{}^{Y}F_{ac}{}^{T}
\Dpartial_bF_{cd}{}^{V}\bar\chi^{Z}\gamma_{d}\,\bar\chi^{W}
+22\,F_{ab}{}^{V}F_{ac}{}^{Y}
\Dpartial_bF_{cd}{}^{T}\bar\chi^{Z}\gamma_{d}\,\bar\chi^{W}
\nonumber\\
&&\qquad -4\,F_{ab}{}^{Y}F_{cd}{}^{T}
\Dpartial_eF_{ac}{}^{V}\bar\chi^{Z}\gamma_{bde}\,\bar\chi^{W}
+4\,F_{ab}{}^{Y}F_{ac}{}^{T}
\Dpartial_cF_{de}{}^{V}\bar\chi^{Z}\gamma_{bde}\,\bar\chi^{W}
\nonumber\\
&&\qquad +4\,F_{ab}{}^{T}F_{cd}{}^{Y}F_{ce}{}^{V}\bar\chi^{Z}\gamma_{abd}\,
\Dpartial_e\chi^{W}
-8\,F_{ab}{}^{Y}F_{cd}{}^{T}F_{ce}{}^{V}\bar\chi^{Z}\gamma_{abd}\,
\Dpartial_e\chi^{W}
\nonumber\\
&&\qquad +6\,F_{ab}{}^{V}F_{cd}{}^{Y}F_{ce}{}^{W}\bar\chi^{Z}\gamma_{abd}\,
\Dpartial_e\chi^{T}
+5\,F_{ab}{}^{V}F_{cd}{}^{W}F_{ce}{}^{Y}\bar\chi^{Z}\gamma_{abd}\,
\Dpartial_e\chi^{T}
\nonumber\\
&&\qquad +6\,F_{ab}{}^{Y}F_{ac}{}^{T}F_{de}{}^{V}\bar\chi^{Z}\gamma_{bcd}\,
\Dpartial_e\chi^{W}
-2\,F_{ab}{}^{Y}F_{ac}{}^{T}F_{de}{}^{Z}\bar\chi^{V}\gamma_{bcd}\,
\Dpartial_e\chi^{W}
\nonumber\\
&&\qquad +4\,F_{ab}{}^{Y}F_{ac}{}^{V}F_{de}{}^{Z}\bar\chi^{W}\gamma_{bcd}\,
\Dpartial_e\chi^{T}
+4\,F_{ab}{}^{T}F_{cd}{}^{V}F_{ce}{}^{Y}\bar\chi^{Z}\gamma_{abd}\,
\Dpartial_e\chi^{W}
\nonumber\\
&&\qquad -4\,F_{ab}{}^{Y}F_{cd}{}^{V}F_{ce}{}^{W}\bar\chi^{Z}\gamma_{abd}\,
\Dpartial_e\chi^{T}
\nonumber\\
&&\qquad +\tfrac{1}{2}\,F_{ab}{}^{Y}F_{cd}{}^{T}F_{ef}{}^{V}\bar\chi^{Z}\gamma_
{abcde}\,
\Dpartial_f\chi^{W}
+\tfrac{1}{2}\,F_{ab}{}^{Y}F_{cd}{}^{T}
   F_{ef}{}^{Z}\bar\chi^{V}\gamma_{abcde}\,
\Dpartial_f\chi^{W}\bigg]\,.
\nonumber
\end{eqnarray}
All authors \cite{kita,RSTZ,KS1,Bilal} agree on the the bosonic terms
 $(\Dpartial F)^2F^2$. Our bosonic terms $F^5$ agree with \cite{KS1}, but are
 given here in a different parametrization. The higher derivative
 terms with fermions agree with \cite{Bilal}.

Note that the group structure is completely specified in terms of
 structure constants. This was not assumed at the start of our calculation.
 In fact,
 the Ansatz was given in terms of traces of four and five
 generators, for which only
 the cyclic property was used. In \cite{Bilal} it is
 shown that all terms with four fields can be written in terms of
 structure constants. We now find that all terms with five fields
 allow such a formulation as well.

The implication of this is that if the group
 contains a $U(1)$ factor, the corresponding
 $U(1)$ fields, which are certainly present at order $\ap{0}$ and $\ap{2}$,
 do not occur in the $\ap{3}$ action. It also implies that the action
 (\ref{L3}) is trivially invariant under the nonlinear supersymmetry
 present at order $\ap{0}$ and $\ap{2}$. The nonlinear transformation
 acts at order $\ap{0}$ only on $\chi$ (at order $\ap{2}$ there are
 modifications \cite{goteborg}) as
\begin{eqnarray}
\label{transNL}
  \delta\chi^A = \eta^A\,,
\end{eqnarray}
 where $\eta$ is a constant spinor, satisfying $f^{ABC}\eta^C=0$. This
 implies that $\eta$ commutes with all group generators, and must
 therefore be in a $U(1)$ factor. The invariance of (\ref{L3})
 under (\ref{transNL}) is then obvious.

The required $\ap{3}$ modifications to the transformation rules for the
 Yang-Mills vector and fermions are
 presented in Appendix \ref{Trans}. We only show supersymmetry
 transformations
 that may modify the supersymmetry algebra with additional field dependent
 gauge transformations. That leaves many supersymmetry transformations that
 are proportional to the lowest order equations of motion. Those will
 modify the on-shell terms in the algebra, but play no role in
 the closure. Since we do not consider quartic fermions in the action
 we cannot say anything about terms bilinear in $\chi$ in the transformation
 rules, nor about closure of the algebra on $\chi$. On $A$ we have
 checked that the algebra closes, and obtain the following new
 gauge transformations in addition to those of order $\ap{0}$ (\ref{algebra})
 and $\ap{2}$ \cite{goteborg}:
\begin{eqnarray}
   {[\d_{\e_1},\d_{\e_2}]} A_a{}^Z &=&
      2\bar{\e_1}{\pd}\slash\e_2
   A_a{}^Z - \D_a \left(2\bar{\e_1}\gamma_b\e_2 A_b{}^Z\right)
\nonumber\\
&& + f^{XYZ}f^{VWX} \Dpartial_a\big(
-16 \Dpartial_{b}F_{cd}{}^V F_{be}{}^{Y} F_{cd}{}^{W}
         \,\bar\epsilon_1\gamma_{e}\epsilon_2%
+8  \Dpartial_{b}F_{cd}{}^V F_{be}{}^{W} F_{cd}{}^{Y}
         \,\bar\epsilon_1\gamma_{e}\epsilon_2%
\nonumber\\
&&\label{algA}
-16  \Dpartial_{b}F_{cd}{}^V F_{be}{}^{W} F_{ce}{}^{Y}
         \,\bar\epsilon_1\gamma_{d}\epsilon_2%
-2  \Dpartial_{b}F_{cd}{}^V F_{ef}{}^{Y} F_{bg}{}^{W}
         \,\bar\epsilon_1\gamma_{cdefg}\epsilon_2 \big)%
\end{eqnarray}

\section{String theory and higher orders in $\ap{}$\label{Zeta}}

In \cite{Bilal} the relation between the tree-level open string
 four-point function and the effective action was explored to
 order $\ap{4}$. In this section we will discuss the relation
 between this four-point function and supersymmetric invariants in
 the effective action, also at higher orders in $\ap{}$. The
 string theory four-point function takes on the following form:
 \begin{eqnarray}
    A_4&=&-8ig^2\, K(1,2,3,4)\,\big(T_{1}^{ABCD}G(s,u) +
     T_{2}^{ABCD}G(s,t) + T_{3}^{ABCD}G(t,u) \big)\,,
 \end{eqnarray}
where
\begin{eqnarray}
   T_1^{ABCD} &=& \Tr\, T^AT^BT^CT^D + \Tr \,T^DT^CT^BT^A\,,
 \nonumber\\
   T_2^{ABCD} &=& \Tr \,T^AT^BT^DT^C + \Tr \,T^CT^DT^BT^A\,,
 \nonumber\\
   T_3^{ABCD} &=& \Tr \,T^AT^CT^BT^D + \Tr \,T^DT^BT^CT^A\,,
\end{eqnarray}
 $g$ is the Yang-Mills coupling constant and $s,t$ and $u$ are the
 standard Mandelstam variables satisfying $s+t+u=0$. $K$ contains the
 polarization and wave-functions of the external lines, where the different
 permutations have to be taken into account.
 The last factor in $A_4$ can always be 
 written as a sum of terms that are proportional to $T_1+T_2+T_3$ 
 (which is the symmetric 
 trace) and $T_i-T_j$ (which can be written in terms of structure constants
 only):
\bea \label{tracestructures}
   &&\tfrac{1}{3}(T_1-T_2) \big( G(s,u)+G(t,u)-2G(s,t) \big) \nn\\
   &&\qquad+ \tfrac{1}{3}(T_1-T_3) \big( G(s,u)+G(s,t)-2G(t,u) \big) \\
   &&\qquad\qquad+\tfrac{1}{3}(T_1+T_3+T_3) 
         \big( G(s,u)+G(t,u)+G(s,t) \big). \nn
\eea
The Veneziano amplitude $G$ contains the $\ap{}$ dependence:
 \begin{eqnarray}
    G(s,t) &=& {1\over st}
    {\Gamma(1-\ap{}s)\Gamma(1-\ap{}t)\over \Gamma(1-\ap{}(s+t))}\,,
 \end{eqnarray}
 and can be expanded in orders of $\ap{}$.

$A_4$ has to be reproduced by the effective action. 
 At order $\ap{0}$ the standard
 Yang-Mills action gives, from the $(A_a{}^A)^4$ vertex and
 from a reducible diagram
 involving three-point vertices, the correct four-point function.
 At higher orders
 in $\ap{}$ it is always the irreducible four-point vertex   
 $\ap{p} \Dpartial^{2p-4} F^4$,
 where the derivatives have to be distributed
 in agreement with the kinematic factors in $A_4$, which yields the
 string four-point function. 
 Therefore we can read off from the string four-point function what
 the coefficients of the terms in the effective action will be.

Using the Taylor expansion for $\log \Gamma(1+z)$,
 \bea
   \log \Gamma(1+z) &=& -\g z + \sum_{m=2}^{\infty}(-1)^m\zeta(m)\frac{z^m}{m},
 \eea
 we obtain the following expression for $G$:
 \bea
   G(s,t) &=& \frac{1}{st}\exp\big\{\sum_{m=2}^{\infty}\ap{m}\frac{\zeta(m)}{m}
              (s^m+t^m-(s+t)^m)\big\}.
 \eea
 $\zeta(n)$  is the Riemann zeta-function, $\g$ 
 the Euler-Mascheroni constant. The expansion of the exponential gives the 
 required result in orders of $\ap{}$, of which the first few terms read:
 \begin{eqnarray}
   G(s,t) &=&  + \ap{0}\,{1\over st}   \nonumber\\
          &&   - \ap{2}\,\tfrac{1}{6}\,\pi^2  \nonumber\\
          &&   - \ap{3}\,(s + t) \zeta(3) \nonumber\\
          &&   - \ap{4}\,\tfrac{1}{360}\, \pi^4 (4 s^2 + s t + 4 t^2)
                   \nonumber\\
          &&   + \ap{5}\, \big( \tfrac{1}{6}\,\pi^2 st(s+t)\zeta(3)
                       -(s+t)(s^2 + st + t^2)\zeta(5) \big) \nonumber\\
          &&   - \ap{6}\,\big( 
      \tfrac{1}{15120}\, \pi^6(16s^4 + 12s^3t + 23s^2t^2 + 12st^3 + 16t^4)
                      - \half\, st(s + t)^2 \zeta(3)^2 \big) \nonumber\\
          &&   + \ap{7}\, \big( 
   \tfrac{1}{360}\,\pi^4 st(s + t)(4s^2 + st + 4t^2)\zeta(3)
                     \nonumber\\
  \label{Veneznab}
          &&\qquad + 
          \tfrac{1}{6}\,\pi^2 st(s^2 + st + t^2)\zeta(5)
                - (s^2 + st + t^2)^2 \zeta(7) \big)\, + \ldots
 \end{eqnarray}
 In this way we understand that the series at even $p$ involving only
 powers of $\pi$ and no $\zeta$-functions corresponds to the supersymmetric 
 invariant that starts  at order $\ap{2}$.
 Similarly, the series of terms with $\zeta(3)^k$ at order
 $p=3k,\ k=1,2\ldots$ is the invariant that starts
 at order $p=3$. We see now that necessarily a new invariant starts
 at every odd power of $\ap{}$.
 For instance, the term with $\zeta(5)$ at order $p=5$
 can only be part of the $p=3$ invariant if there were
 a relation with rational
 coefficients between $\pi^2\zeta(3)$ and $\zeta(5)$.
 To our knowledge, no such relation
 between the $\zeta(2n+1)$ for different $n$ exist,
 and new invariants therefore
 appear at all odd orders of $\ap{}$.

The leading term with $\ap{n}\zeta(n)$ is
 proportional to  $(s^n+t^n-(s+t)^n)/st$. 
 For $n$ {\it odd} 
 this is of the form
 \begin{equation}
   (s+t)P(s,t) {\rm ~~, with~~~} P(s,t) = -\frac{s^n+t^n+u^n}{stu}.
 \end{equation}
Now in (\ref{tracestructures})
 the symmetric trace is proportional to $G(s,t)+G(s,u)+G(t,u)$, 
 which for the leading term
 with $\ap{n}\zeta(n)$, $n$ odd, gives a factor:
\begin{equation}
   (s+t)P(s,t)+(s+u)P(s,u)+(t+u)P(t,u)=2(s+t+u)P(s,t)=0.
\end{equation}
 Therefore, all new invariants starting at $\ap{n}$ for $n$ odd 
 can be expressed in terms of structure 
 constants only, and thus vanish in the abelian limit.

The conclusion must be that supersymmetry
 by itself cannot be sufficient to determine
 the open string effective action. The effective action is a sum of an
 infinite number of superinvariants, of which the relative coefficients can be
 determined from string theory,
 but not from supersymmetry alone. Our
 argument does not exclude the possiblity that additional invariants,
 which do not contribute to the four-point function, appear in
 the effective action.

In the abelian case $A_4$ simplifies to
\begin{equation}
   A_4 = -8ig^2\, K(1,2,3,4)\,\big(G(s,u) +
    G(s,t) + G(t,u) \big)\,.
\end{equation}
The expansion in $\ap{}$ now reads
\begin{eqnarray}
 && G(s,u)
   + G(s,t) + G(t,u) = \nonumber\\
 &&\qquad -\ap{2}\, \half\,\pi^2 \nonumber\\
 &&\qquad -\ap{4}\, \tfrac{1}{24}\,\pi^4 (s^2 + st + t^2) \nonumber\\
 &&\qquad +\ap{5}\, \half\,\pi^2 st(s + t)\zeta(3) \nonumber\\
 &&\qquad -\ap{6}\, \tfrac{1}{240}\,\pi^6 (s^2 + st + t^2)^2
   \nonumber\\
\label{Venezabel}
 &&\qquad +\ap{7}\, \tfrac{1}{48}\,\pi^2st(s^3 + 2s^2t + 2st^2 + t^3)
         \big(2\pi^2\zeta(3) + 24\zeta(5)\big) + \ldots
\end{eqnarray}
where we have used $s+t+u=0$. Of course there is now no order
 $\ap{0}$ term, also the term at order $\ap{3}$ vanishes. However,
 at order $\ap{4}$ there is a four-point function, which in
 the effective action must be represented by a term $\ap{4}\partial^{4} F^4$.
 Such terms can indeed be found in the analysis of \cite{AndTs,Wyll}.
 Since the terms at order $\ap{4}$ without derivatives on $F$ belong to
 the Born-Infeld superinvariant, these higher derivatives must
 be invariant by themselves.

The expansion (\ref{Venezabel}) shows  terms
 proportional to $\pi^2\zeta(2k+1)$ at
 odd orders  $\ap{2k+3}$. These also appear in the expansion
 of $G(s,t)$ that we presented for the nonabelian case (\ref{Veneznab}).  
 There it would be tempting to interpret these terms as an
 ``interference'' between the $\ap{2}$ invariant and the $\ap{2k+1}$
 invariant proportional to $\zeta(2k+1)$. In that case they
 would be required to cancel the $\ap{2}$ variation of the 
 $\zeta(2k+1)$-invariant and the $\ap{2k+1}$ variation
 of the $\ap{2}$ invariant. However, if that interpretation were correct,
 these terms should vanish in the abelian case, because the
 $\ap{2k+1}$ invariant does. A closer look at (\ref{tracestructures})
 shows that in the nonabelian case these terms contain only
 the symmetric trace contribution, and not the terms $T_i-T_j$, 
 proportional to structure constants. Therefore, they
 correspond to independent invariants in the nonabelian case,
 which survive the abelian limit.

\section{Discussion\label{Discussion}}

In this paper we have obtained the contribution to the open superstring
 effective action at order $\ap{3}$, with the exception of terms
 quartic in the fermions. We assume that the nonabelian structure
 is given by a single trace of group generators, in agreement
 with what one would expect from tree level string theory. 
 The result is then
 unique, up to total derivatives and field redefinitions.
 In the sectors that allow comparison
 with previous work we agree with  \cite{KS1,Bilal}.
 We disagree with the result of \cite{RSTZ}, which is an
 effective action in four dimensions. That does not imply that the
 action of \cite{RSTZ} is not supersymmetric - it may well be that
 more invariants can be found in four than in ten dimensions.

The traces over group generators turn out to give products of structure
 constants only. It was known that the nonabelian result should vanish
 in the abelian limit, but that is a much weaker statement than structure
 constants only. It implies that fields in $U(1)$ factors of the gauge
 group are absent from this part of the effective action, and that therefore
 the nonlinear supersymmetry is trivial.

Although our procedure works for an arbitrary gauge group at order $\ap{3}$,
 we do not expect this to hold at higher orders. 
 Continuing the
 iteration to order $\ap{4}$ would give two kinds of contributions. 
 In the first
 place there are terms that come from the variation of the 
 order $\ap{4}$ Ansatz
 with the $\ap{0}$ transformation rules. If we still assume the Ansatz to be
 proportional to a single trace, these terms are proportional 
 to $\Tr(T^AT^BT^CT^DT^ET^F)$.
 In the second place there are contributions from the variation of 
 the $\ap{2}$ action
 with the $\ap{2}$ transformation rules. Such terms are proportional 
 to a product of 
 two traces, $\Tr(T^AT^BT^CT^G)\Tr(T^DT^ET^FT^G)$. These different terms
 can only communicate with each other if the generators $T$ satisfy
 requirements which are analogous to the unitarity conditions
 on Chan-Paton factors. We therefore expect that at higher
 orders supersymmetry requires the generators to be 
 in the fundamental
 representation of $U(n)$, $SO(n)$ or $USp(n)$. 

We have argued that the $\ap{3}$ invariant is just the first of an
 infinite number of invariants appearing at all odd orders $\ap{2p+1}$
 in the effective action, with a coefficient proportional to
 $\zeta(2p+1)$. This sheds new light on efforts \cite{BdRS} to obtain
 the effective action through $\kappa$-symmetry, a method, which was
 extremely successfull in the abelian situation. $\kappa$-symmetry with
 parameters in the adjoint representation of the gauge group turns out
 not to work \cite{BBRS}, $\kappa$-symmetry that only transforms
 fields in the $U(1)$ direction of the group will not see the $\ap{3}$
 effective action we have just obtained. The most likely scenario, if
 $\kappa$-symmetry works at all in the nonabelian context, is that
 it gives the part of the action generated by the $\ap{2}$ terms, i.e.,
 the terms that are not proportional to $\zeta$-functions.

In \cite{KS2} the result of \cite{KS1} was tested by calculating the
 spectrum of the deformed Yang-Mills theory in a constant
 magnetic background. By T-duality the constant magnetic field
 corresponds to D-branes at angles, and in this context string theory
 allows an alternative calculation of the spectrum
 \cite{wati}. This test uses
 configurations of Yang-Mills fields in the Cartan subalgebra of
 the gauge group. It would be interesting to find a true nonabelian
 generalization of the method of \cite{wati}, also including fermions.

Terms with derivatives in the field strength $F$ are inevitably
 present in the nonabelian effective action, and also in the
 abelian case there is no reason to assume that such terms are small
 in general. In Section 4 we have discussed such terms in the context
 of the open string four-point function. From the plethora of
 supersymmetric invariants that are indicated by the
 four-point function, it is clear that the construction of the
 complete open string effective action, in both the abelian and the nonabelian
 cases,  requires perhaps additional symmetries beyond supersymmetry,
 but certainly new
 insights. One may conclude that the real surprise in this field is
 still the apparent simplicity of the abelian Born-Infeld action,
 which disappears completely as soon as one deviates from the
 context of slowly varying abelian fields.

\section*{Acknowledgements}
\noindent
 We thank Eric Bergshoeff and Paul Koerber for useful discussions.
 This work is supported by the European Commission
 RTN programme HPRN-CT-2000-00131, in which we are associated
 to the university of Utrecht. The work of M.G.C.E.\ is part of
 the research programme of the Stichting FOM, Utrecht.

\appendix

\section{Conventions\label{Conventions}}

We consider a compact gauge group $G$ and parametrize elements $g$
 that are connected to the identity by $g=\exp\,\Lambda\cdot T$.
 The generators $T^A$ satisfy the orthonormality
 condition $\Tr\,T^AT^B=-\d^{AB}$ and the algebra
\begin{equation}
  [T^A,T^B]=f^{ABC}T^C,
\end{equation}
 where the $f^{ABC}$ are completely antisymmetric real structure
 constants. No further restrictions are imposed on the generators.
 We freely raise and lower indices on the structure constants. All
 fields in this paper transform in the adjoint
 representation, $(T^A_{\rm adj})^{BC}=-f^{ABC}$. For such fields
 we use the notation $\Phi\equiv \Phi\cdot T = \Phi^A T^A$,
 where $T^A$ can be any representation. Since under a gauge
 transformation $\Phi\rightarrow g\Phi g^{-1}$, we can form gauge
 invariant objects by tracing, e.g. $\Tr\, \Phi_1\cdots\Phi_n$.
\\
 The infinitesimal gauge transformations of the nonabelian Yang-Mills
 multiplet $(A_a,\chi)$ are
\bea
  \d A_a     &=&  -\D_a \Lambda, \\
  \d F_{ab}  &=&  [\Lambda,F_{ab}], \\
  \d \chi    &=&  [\Lambda,\chi].
\eea
The covariant derivative and the field strength are defined by
\bea
  \D_a \Phi          &=&  \pd_a\Phi + [A_a,\Phi],\\
  {[\D_a,\D_b]}\Phi  &=&  [F_{ab},\Phi],
\eea
so that
\begin{equation}
  F_{ab} \;=\; \pd_a A_b - \pd_b A_a + [A_a, A_b].
\end{equation}
$F$ satisfies a Bianchi identity: $\D_{[a} F_{bc]}\equiv 0$.

\section{Transformation rules\label{Trans}}

We now present the supersymmetry
 transformation rules that leave the action (\ref{L3})
 invariant. The
 transformation rules of the fermions are:

\renewcommand{\arraystretch}{1.3}
\begin{eqnarray}
\delta_3\chi^Z &=&
 f^{XYZ} f^{VWX}\bigg[\,
 -4\,\Dpartial_{a}F_{bc}{}^Y\Dpartial_{b}F_{ad}{}^VF_{cd}{}^W
 +2 \,\Dpartial_{a}F_{bc}{}^Y\Dpartial_{a}F_{bd}{}^VF_{cd}{}^W 
\nonumber\\
&&
+4 \,\Dpartial_{a}F_{bc}{}^Y\Dpartial_{a}F_{bd}{}^VF_{ce}{}^W\gamma_{de}%
-6 \,\Dpartial_{a}F_{bc}{}^V\Dpartial_{a}F_{bd}{}^YF_{ce}{}^W\gamma_{de}%
\nonumber\\
&&
-2 \,\Dpartial_{a}F_{bc}{}^Y\Dpartial_{b}F_{ad}{}^VF_{ce}{}^W\gamma_{de}%
-2 \,\Dpartial_{a}F_{bc}{}^Y\Dpartial_{b}F_{de}{}^VF_{ad}{}^W\gamma_{ce}%
\nonumber\\
&&
+2 \,\Dpartial_{a}F_{bc}{}^Y\Dpartial_{d}F_{be}{}^VF_{ad}{}^W\gamma_{ce}%
+2 \,\Dpartial_{a}F_{bc}{}^Y\Dpartial_{a}F_{de}{}^VF_{bd}{}^W\gamma_{ce}%
\nonumber\\
&&
-3 \,\Dpartial_{a}F_{bc}{}^Y\Dpartial_{a}F_{de}{}^VF_{bc}{}^W\gamma_{de}%
+\tfrac{3}{2}\,
     \Dpartial_{a}F_{bc}{}^V\Dpartial_{a}F_{de}{}^YF_{bc}{}^W\gamma_{de}%
\nonumber\\
&&
+\tfrac{3}{2}\,
     \Dpartial_{a}F_{bc}{}^Y\Dpartial_{a}F_{bc}{}^VF_{de}{}^W\gamma_{de}%
- \,\Dpartial_{a}\Dpartial_{b}F_{cd}{}^{Y}F_{be}{}^VF_{cd}{}^W\gamma_{ae}%
\nonumber\\
&&
-4 \,\Dpartial_{a}\Dpartial_{b}F_{cd}{}^{V}F_{ae}{}^YF_{be}{}^W\gamma_{cd}%
+3 \,\Dpartial_{a}\Dpartial_{b}F_{cd}{}^{V}F_{ae}{}^WF_{be}{}^Y\gamma_{cd}%
\nonumber\\
&&
-3 \,\Dpartial_{a}F_{bc}{}^Y\Dpartial_{a}F_{de}{}^VF_{df}{}^W\gamma_{bcef}%
- \,\Dpartial_{a}F_{bc}{}^V\Dpartial_{a}F_{de}{}^YF_{df}{}^W\gamma_{bcef}%
\nonumber\\
&&
+3 \,\Dpartial_{a}F_{bc}{}^Y\Dpartial_{a}F_{bd}{}^VF_{ef}{}^W\gamma_{cdef}%
- \,\Dpartial_{a}F_{bc}{}^Y\Dpartial_{d}F_{ef}{}^VF_{ad}{}^W\gamma_{bcef}%
\nonumber\\
&& \label{transchi}
+\tfrac{1}{4} \,\Dpartial_{a}F_{bc}{}^Y\Dpartial_{a}
F_{de}{}^VF_{fg}{}^W\gamma_{bcdefg}\bigg]\,\epsilon%
\ +
 \\
&&
+ f^{WXY} f^{TUV} f^{TWZ}
     \bigg[7\,F_{ab}{}^XF_{ac}{}^UF_{de}{}^VF_{de}{}^Y\gamma_{bc}%
\nonumber\\
&&
-2 \,F_{ab}{}^XF_{cd}{}^UF_{ae}{}^VF_{ce}{}^Y\gamma_{bd}%
-6 \,F_{ab}{}^XF_{cd}{}^UF_{ae}{}^YF_{ce}{}^V\gamma_{bd}%
\nonumber\\
&&
-4 \,F_{ab}{}^XF_{cd}{}^UF_{ef}{}^YF_{ce}{}^V\gamma_{abdf}%
-\tfrac{3}{2} \,F_{ab}{}^XF_{cd}{}^U
F_{ef}{}^VF_{eg}{}^Y\gamma_{abcdfg}\bigg]\,\epsilon%
\ + \nonumber\\
&&
+ f^{XYZ} f^{UVW} f^{TUX}
   \bigg[2\,F_{ab}{}^YF_{ac}{}^VF_{de}{}^WF_{de}{}^T\gamma_{bc}%
\nonumber\\
&&
+2 \,F_{ab}{}^VF_{cd}{}^WF_{ae}{}^YF_{ce}{}^T\gamma_{bd}%
-8 \,F_{ab}{}^VF_{cd}{}^TF_{ae}{}^YF_{ce}{}^W\gamma_{bd}%
\nonumber\\
&&
-4 \,F_{ab}{}^YF_{cd}{}^VF_{ef}{}^TF_{ce}{}^W\gamma_{abdf}%
+ \,F_{ab}{}^YF_{cd}{}^VF_{ef}{}^WF_{ef}{}^T\gamma_{abcd}%
\nonumber\\
&&
+ \,F_{ab}{}^VF_{cd}{}^TF_{ef}{}^WF_{eg}{}^Y\gamma_{abcdfg}\bigg]\,\epsilon%
\,.\nonumber
\end{eqnarray}
\renewcommand{\arraystretch}{1}
\noindent The transformation rules for the vector field are:
\renewcommand{\arraystretch}{1.3}
\begin{eqnarray}
\delta_3 A_a^{Z} &=&
 f^{XYZ} f^{VWX}\bigg[
 +2\,\Dpartial_{b}F_{cd}{}^{Y}\Dpartial_{b}F_{cd}{}^{W}
  \,\bar\epsilon\,\gamma_{a}\chi^{V}%
 -\,\Dpartial_{a}\Dpartial_{b}F_{cd}{}^{W}F_{cd}{}^{Y}
  \,\bar\epsilon\,\gamma_{b}\chi^{V}%
\nonumber\\
&&
 +5\,\Dpartial_{a}\Dpartial_{b}F_{cd}{}^{W}F_{cd}{}^{V}
  \,\bar\epsilon\,\gamma_{b}\chi^{Y}%
 +4\,\Dpartial_{b}F_{ac}{}^{Y}\Dpartial_{b}F_{cd}{}^{W}
  \,\bar\epsilon\,\gamma_{d}\chi^{V}%
\nonumber\\
&&
 -\,\Dpartial_{a}F_{bc}{}^{Y}\Dpartial_{d}F_{bc}{}^{W}
  \,\bar\epsilon\,\gamma_{d}\chi^{V}%
 -5\,\Dpartial_{a}F_{bc}{}^{W}\Dpartial_{d}F_{bc}{}^{V}
  \,\bar\epsilon\,\gamma_{d}\chi^{Y}%
\nonumber\\
&&
 +2\Dpartial_{b}F_{ac}{}^{W}\Dpartial_{b}F_{de}{}^{Y}
  \,\bar\epsilon\,\gamma_{cde}\chi^{V}%
 +\,\Dpartial_{b}F_{cd}{}^{Y}\Dpartial_{b}F_{ef}{}^{W}
  \,\bar\epsilon\,\gamma_{acdef}\chi^{V}%
\nonumber\\
&&
 +2\,\Dpartial_{b}F_{cd}{}^{Y}F_{cd}{}^{W}
  \,\bar\epsilon\,\gamma_{a}\Dpartial_{b}\chi^{V}%
 +2\,\Dpartial_{b}F_{cd}{}^{W}F_{cd}{}^{Y}
  \,\bar\epsilon\,\gamma_{a}\Dpartial_{b}\chi^{V}%
\nonumber\\
&&
 -2\,\Dpartial_{b}F_{cd}{}^{Y}F_{ac}{}^{W}
  \,\bar\epsilon\,\gamma_{b}\Dpartial_{d}\chi^{V}%
 -2\,\Dpartial_{b}F_{cd}{}^{W}F_{cd}{}^{Y}
  \,\bar\epsilon\,\gamma_{b}\Dpartial_{a}\chi^{V}%
\nonumber\\
&&
 +6\,\Dpartial_{b}F_{cd}{}^{W}F_{cd}{}^{V}
  \,\bar\epsilon\,\gamma_{b}\Dpartial_{a}\chi^{Y}%
 -2\,\Dpartial_{b}F_{ac}{}^{Y}F_{bd}{}^{W}
  \,\bar\epsilon\,\gamma_{c}\Dpartial_{d}\chi^{V}%
\nonumber\\
&&
 +4\,\Dpartial_{b}F_{ac}{}^{W}F_{bd}{}^{Y}
  \,\bar\epsilon\,\gamma_{c}\Dpartial_{d}\chi^{V}%
 -8\,\Dpartial_{a}F_{bc}{}^{Y}F_{bd}{}^{W}
  \,\bar\epsilon\,\gamma_{c}\Dpartial_{d}\chi^{V}%
\nonumber\\
&&
 -2\,\Dpartial_{a}F_{bc}{}^{W}F_{bd}{}^{V}
  \,\bar\epsilon\,\gamma_{c}\Dpartial_{d}\chi^{Y}%
 +4\,\Dpartial_{b}F_{cd}{}^{W}F_{ac}{}^{Y}
  \,\bar\epsilon\,\gamma_{d}\Dpartial_{b}\chi^{V}%
\nonumber\\
&&
 -10\,\Dpartial_{a}F_{bc}{}^{W}F_{bd}{}^{Y}
  \,\bar\epsilon\,\gamma_{d}\Dpartial_{c}\chi^{V}%
 +2\,\Dpartial_{a}F_{bc}{}^{W}F_{bd}{}^{V}
  \,\bar\epsilon\,\gamma_{d}\Dpartial_{c}\chi^{Y}%
\nonumber
\end{eqnarray}
\begin{eqnarray}
&&
 +2\,\Dpartial_{b}F_{cd}{}^{Y}F_{be}{}^{W}
  \,\bar\epsilon\,\gamma_{acd}\Dpartial_{e}\chi^{V}%
 +2\,\Dpartial_{b}F_{cd}{}^{Y}F_{ce}{}^{W}
  \,\bar\epsilon\,\gamma_{ade}\Dpartial_{b}\chi^{V}%
\nonumber\\
&&
 +2\,\Dpartial_{b}F_{ac}{}^{W}F_{de}{}^{Y}
  \,\bar\epsilon\,\gamma_{bde}\Dpartial_{c}\chi^{V}%
 -2\,\Dpartial_{a}F_{bc}{}^{W}F_{de}{}^{Y}
  \,\bar\epsilon\,\gamma_{bce}\Dpartial_{d}\chi^{V}%
\nonumber\\
&&
 +2\,\Dpartial_{a}F_{bc}{}^{W}F_{de}{}^{V}
  \,\bar\epsilon\,\gamma_{bce}\Dpartial_{d}\chi^{Y}%
 +\,\Dpartial_{b}F_{ac}{}^{Y}F_{de}{}^{W}
  \,\bar\epsilon\,\gamma_{cde}\Dpartial_{b}\chi^{V}%
\nonumber\\
&&
 -2\,\Dpartial_{b}F_{ac}{}^{W}F_{de}{}^{Y}
  \,\bar\epsilon\,\gamma_{cde}\Dpartial_{b}\chi^{V}%
 +\,\Dpartial_{b}F_{cd}{}^{Y}F_{ae}{}^{W}
  \,\bar\epsilon\,\gamma_{cde}\Dpartial_{b}\chi^{V}%
\nonumber\\
&&
 -\tfrac{1}{2}\,\Dpartial_{b}F_{cd}{}^{Y}F_{ef}{}^{W}
  \,\bar\epsilon\,\gamma_{acdef}\Dpartial_{b}\chi^{V}%
 +\,\Dpartial_{b}F_{cd}{}^{W}F_{ef}{}^{Y}
  \,\bar\epsilon\,\gamma_{acdef}\Dpartial_{b}\chi^{V}%
\nonumber\\
&&
 +10\,F_{bc}{}^{Y}F_{bd}{}^{W}
  \,\bar\epsilon\,\gamma_{a}\Dpartial_{c}\Dpartial_{d}\chi^{V}%
 -8\,F_{bc}{}^{W}F_{bd}{}^{Y}
  \,\bar\epsilon\,\gamma_{a}\Dpartial_{c}\Dpartial_{d}\chi^{V}%
\nonumber\\
&&
 -2\,F_{ab}{}^{Y}F_{cd}{}^{W}
  \,\bar\epsilon\,\gamma_{d}\Dpartial_{b}\Dpartial_{c}\chi^{V}%
 +2\,F_{ab}{}^{W}F_{cd}{}^{Y}
  \,\bar\epsilon\,\gamma_{d}\Dpartial_{b}\Dpartial_{c}\chi^{V}%
\nonumber\\
&&
 -8\,F_{bc}{}^{W}F_{bd}{}^{Y}
  \,\bar\epsilon\,\gamma_{d}\Dpartial_{c}\Dpartial_{a}\chi^{V}%
 -2\,F_{bc}{}^{Y}F_{de}{}^{W}
  \,\bar\epsilon\,\gamma_{cde}\Dpartial_{a}\Dpartial_{b}\chi^{V}%
\nonumber\\
&&
 -2\,F_{bc}{}^{W}F_{de}{}^{V}
  \,\bar\epsilon\,\gamma_{cde}\Dpartial_{a}\Dpartial_{b}\chi^{Y}\bigg]
\ +\nonumber\\
&& \label{transA}
+ f^{XYZ} f^{UVW} f^{TUX}\bigg[ +13\,F_{ab}{}^{W}F_{cd}{}^{Y}F_{cd}{}^{V}
  \,\bar\epsilon\,\gamma_{b}\chi^{T}%
\\
&&
 -3\,F_{ab}{}^{W}F_{cd}{}^{Y}F_{cd}{}^{T}
  \,\bar\epsilon\,\gamma_{b}\chi^{V}%
 -5\,F_{ab}{}^{W}F_{cd}{}^{V}F_{cd}{}^{T}
  \,\bar\epsilon\,\gamma_{b}\chi^{Y}%
\nonumber\\
&&
 -20\,F_{bc}{}^{W}F_{ad}{}^{Y}F_{bd}{}^{V}
  \,\bar\epsilon\,\gamma_{c}\chi^{T}%
 +22\,F_{bc}{}^{W}F_{ad}{}^{T}F_{bd}{}^{V}
  \,\bar\epsilon\,\gamma_{c}\chi^{Y}%
\nonumber\\
&&
 -6\,F_{bc}{}^{T}F_{ad}{}^{Y}F_{bd}{}^{W}
  \,\bar\epsilon\,\gamma_{c}\chi^{V}%
 +34\,F_{bc}{}^{W}F_{ad}{}^{T}F_{bd}{}^{Y}
  \,\bar\epsilon\,\gamma_{c}\chi^{V}%
\nonumber\\
&&
 +2\,F_{bc}{}^{T}F_{de}{}^{Y}F_{bd}{}^{W}
  \,\bar\epsilon\,\gamma_{ace}\chi^{V}%
 -4\,F_{bc}{}^{Y}F_{de}{}^{W}F_{bd}{}^{V}
  \,\bar\epsilon\,\gamma_{ace}\chi^{T}%
\nonumber\\
&&
 +4\,F_{bc}{}^{W}F_{de}{}^{Y}F_{de}{}^{V}
  \,\bar\epsilon\,\gamma_{abc}\chi^{T}%
 -\tfrac{7}{2}\,F_{bc}{}^{W}F_{de}{}^{Y}F_{de}{}^{T}
  \,\bar\epsilon\,\gamma_{abc}\chi^{V}%
\nonumber\\
&&
 -\,F_{bc}{}^{W}F_{de}{}^{V}F_{de}{}^{T}
  \,\bar\epsilon\,\gamma_{abc}\chi^{Y}%
 +42\,F_{ab}^{W}F_{cd}{}^{Y}F_{ce}{}^{V}
  \,\bar\epsilon\,\gamma_{bde}\chi^{T}%
\nonumber\\
&&
 +6\,F_{ab}^{T}F_{cd}{}^{Y}F_{ce}{}^{W}
  \,\bar\epsilon\,\gamma_{bde}\chi^{V}%
 +\tfrac{1}{2}\,F_{cd}{}^{Y}F_{ef}{}^{V}
  \,\bar\epsilon\,\gamma_{bcdef}\chi^{T}%
\nonumber\\
&&
 +\tfrac{7}{2}\, F_{ab}{}^{W}F_{cd}{}^{Y}F_{ef}{}^{T}
  \,\bar\epsilon\,\gamma_{bcdef}\chi^{V}%
 -\tfrac{5}{2}\, F_{ab}{}^{W}F_{cd}{}^{V}F_{ef}{}^{T}
  \,\bar\epsilon\,\gamma_{bcdef}\chi^{Y}%
\nonumber\\
&&
 +3\,F_{bc}{}^{W}F_{de}{}^{Y}F_{ad}{}^{V}
  \,\bar\epsilon\,\gamma_{bce}\chi^{T}%
 -6\,F_{bc}{}^{W}F_{de}{}^{V}F_{ad}{}^{T}
  \,\bar\epsilon\,\gamma_{bce}\chi^{Y}%
\nonumber\\
&&
 +2\,F_{bc}{}^{W}F_{de}{}^{T}F_{ad}{}^{V}
  \,\bar\epsilon\,\gamma_{bce}\chi^{Y}%
 +\,F_{bc}{}^{T}F_{de}{}^{Y}F_{ad}{}^{W}
  \,\bar\epsilon\,\gamma_{bce}\chi^{V}%
\nonumber\\
&&
 -3\,F_{bc}{}^{W}F_{de}{}^{T}F_{ad}{}^{Y}
  \,\bar\epsilon\,\gamma_{bce}\chi^{V}%
 -5\,F_{bc}{}^{W}F_{de}{}^{Y}F_{df}{}^{V}
  \,\bar\epsilon\,\gamma_{abcef}\chi^{T}%
\nonumber\\
&&
 -\,F_{bc}{}^{T}F_{de}{}^{Y}F_{df}{}^{W}
  \,\bar\epsilon\,\gamma_{abcef}\chi^{V}%
 +\tfrac{1}{4}\,F_{bc}{}^{Y}F_{de}{}^{W}F_{fg}{}^{T}
  \,\bar\epsilon\,\gamma_{abcdefg}\chi^{V}\bigg]
\ +\nonumber\\
&&
 + f^{WXY} f^{TUV} f^{TWZ}\bigg[ +8\,F_{bc}{}^{Y}F_{ad}{}^{X}F_{bd}{}^{V}
  \,\bar\epsilon\,\gamma_{c}\chi^{U}%
\nonumber\\
&&
 +2\,F_{bc}{}^{Y}F_{de}{}^{V}F_{bd}{}^{X}
  \,\bar\epsilon\,\gamma_{ace}\chi^{U}%
 -14\,F_{ab}{}^{Y}F_{cd}{}^{X}F_{ce}{}^{V}
  \,\bar\epsilon\,\gamma_{bde}\chi^{U}%
\nonumber\\
&&
 -4\,F_{bc}{}^{Y}F_{de}{}^{X}F_{ad}{}^{V}
  \,\bar\epsilon\,\gamma_{bce}\chi^{U}%
 +7\,F_{bc}{}^{Y}F_{de}{}^{X}F_{df}{}^{V}
  \,\bar\epsilon\,\gamma_{abcef}\chi^{U}\bigg]\,.%
\nonumber
\end{eqnarray}
\renewcommand{\arraystretch}{1}

\end{document}